\documentclass[final,review]{elsarticle}

\usepackage{url}
\usepackage{float}
\usepackage{graphicx}
\usepackage[table,xcdraw]{xcolor}
\usepackage{caption}
\usepackage{booktabs}
\usepackage{balance}
\usepackage{siunitx}
\DeclareSIUnit\bar{bar}
\usepackage[version=4]{mhchem}
\usepackage[normalem]{ulem}
\usepackage[utf8]{inputenc}  

\biboptions{numbers,sort&compress} 
\usepackage{lineno} 
\usepackage{amssymb,amsmath} 
\usepackage{color} 
\biboptions{numbers,sort&compress} 
\usepackage[hidelinks]{hyperref}
\usepackage[flushleft]{threeparttable} 
\modulolinenumbers[5]
\graphicspath{{./Figures/}} 
\usepackage{soul}	
\soulregister\cite7 
\usepackage[a4paper, total={5in, 8in}]{geometry}

\journal{}

\begin{document}
\begin{frontmatter}

\title{Characterizing and visualizing the direct injection of hydrogen into high-pressure argon and nitrogen environments}

\author{Max Peters\corref{correspondingauthor}}
\cortext[correspondingauthor]{Corresponding author}
\ead{m.e.e.peters@tue.nl}
\author{Noud Maes} 
\author{Nico Dam}
\author{Jeroen van Oijen}
\address{Department of Mechanical Engineering, Eindhoven University of Technology, P.O. Box 513, 5600 MB Eindhoven, the Netherlands}

\begin{abstract}
Using argon as the working fluid in an internal combustion engine holds the potential of substantially enhancing thermal efficiency because of its high specific heat ratio. Burning hydrogen, not with air but with pure oxygen, in a closed-loop argon power cycle would lead to an emission-free exhaust composition, effectively containing only water and argon, which can be separated by condensation. One of the current challenges is the high-pressure direct injection of both fuel and oxidizer, which directly controls the combustion process. To support the development of such an injection strategy, high-pressure injections of hydrogen in pressurized argon and nitrogen are investigated using high-speed Schlieren and pressure transducers in a non-heated constant-volume chamber at varying conditions. We show that hydrogen mass flow is well described by choked flow theory, while jet penetration and angle are determined by the pressure ratio between the fuel line and the ambient. A correlation is presented between jet penetration and pressure ratio for argon or nitrogen ambient at room temperature.

\end{abstract}

\end{frontmatter}

\section{Introduction}
For the increasing energy demand with a simultaneous push toward cleaner technologies, hydrogen is considered as a future energy carrier in a multitude of applications \cite{h2energysystem}. For one of these applications, internal combustion engines (ICE), many different pathways regarding hydrogen injection, ignition, and combustion strategy are considered to optimize power density, emissions, and efficiency \cite{h2enginesreview,xanderSAE}. One approach for pollutant-free, cost-effective power production could be the use of a hydrogen-based Argon Power Cycle (APC): an ICE with argon as the working fluid, in which hydrogen is burned not with air but with pure oxygen. This idea was first patented and introduced by Laumann and Reynolds in 1978 \cite{laumann1978}, followed by a demonstration by de Boer and Hulet on the versatile Combustion Fuel Research (CFR) engine two years later, showing substantial gains in thermal efficiency with high concentrations of argon  \cite{deboer1980}. This ultra-high conversion efficiency power generation cycle is based on the increased thermal efficiency due to an increased specific heat ratio ($\gamma$) of the working fluid. Since heat engines typically use air as the working fluid, the specific heat ratio of the working fluid is dominated by diatomic molecules, leading to $\gamma \approx 1.4$. Working fluids with the highest possible specific heat ratio are monoatomic noble gases such as helium, neon, and argon where $\gamma \approx 1.66$. It makes sense to focus on argon as it is affordable and abundant in the atmosphere. For practical purposes the argon would have to be distilled from liquified air \cite{argondistil}, meaning that it is ideally recirculated in a closed loop configuration for the proposed cycle. As the system does not contain any nitrogen, combustion of hydrogen is not limited by the generation of nitrogen oxide emissions, regardless of temperature \cite{gomes_nox_T}. This allows for a highly efficient, pollution-free power system, which is ideal for use in a hydrogen energy storage scheme \cite{hydrogenstorage}. Because of the closed loop, oxygen needs to be provided to the combustion chamber as well, where the oxygen and hydrogen could be produced by the electrolysis of water in stoichiometric proportions.

Spark ignition (SI) is the easiest way of realizing hydrogen combustion in an ICE, but only at relatively low compression ratios. Kuroki et al.\ \cite{kuroki2010} and Killingsworth et al.\ \cite{killingsworth2011}, showed an increase of thermal efficiency, but both studies were severely limited by knock at compression ratios above 5.5. To optimize efficiency on the longer term, the compression-ignition (CI) mode is desired because of the inherent increase of thermal efficiency along with compression ratio \cite{heywood1988internal}. While this would usually suffer from an increase in nitrogen oxide emissions because of elevated temperatures, substitution of the air with argon would remove this constraint, as indicated above. This introduces the need for high-pressure direct injection of hydrogen and/or oxygen, as both fuel and oxidizer need to mix and react in a controlled manner. In comparison to conventional fuels used in SI or CI concepts, hydrogen will be gaseous during the injection, unless it is extremely cooled. Injecting pressurized gas instead of incompressible liquids has the implication that compressibility effects will have to be considered \cite{oueletteandhill}.

The experimental groundwork on compressed gaseous injections was performed by Ewan and Moodie in 1986 \cite{ewanmoodie}, while Franquet et al.\ \cite{FRANQUET201525} showed in detail that a relatively wide shock barrel and Mach disk (when compared to the nozzle exit hole) will form near the nozzle that could be influenced by nozzle geometry. Ouelette and Tsujimura et al.\ studied jet penetration and introduced semi-empirical models for jet penetration of high-pressure gaseous injections into air \cite{Ouellette_1996,tsujimura}. Mansor et al.\ studied jet penetration and ignition characteristics of high-pressure hydrogen injections into an argon-oxygen atmosphere \cite{mansor}, showing feasibility for auto-igniting jets in a constant-volume vessel at ambient temperatures above 1000 K, but rapid premixed combustion after a long ignition delay below 1000 K. The measurements of Naber and Siebers \cite{NABER1998363} already confirmed the need for higher auto-ignition temperature compared to conventional diesel combustion, but also show a rise in nitrogen oxide emissions when air is used as a working fluid. A recent contribution by Yip et al.\ introduced the geometry and usage of a capped GDI injector for hydrogen injections into high-pressure and high-temperature nitrogen/air environments, with similar auto-ignition results \cite{yip}. Their parametric auto-ignition study revealed that the hydrogen jet would always ignite from the mid-jet region \cite{YIP_para}. In order to further investigate the governing mechanisms for hydrogen jet flame evolution and stabilisation, they ignited the jet using a laser-induced plasma in a more recent study \cite{YIP_laser}. An investigation into this mid-jet region mixing field was recently performed by Wu et al.\ \cite{kaust_raman_rayleigh}.

In order to reach the elevated temperature for auto-ignition, high compression ratios are needed. This research will therefore focus on the lower (2-10) pressure ratios between fuel line and chamber (nPR), but above the choked flow limit of 1.899 for hydrogen, as fuel pressure is expected to be a practical limitation in the future. To aid the understanding of hydrogen jet behavior in an argon environment, to support the development of the pursued argon power cycle, and to aid simultaneous model development, hydrogen injection experiments are performed in a constant-volume chamber. Combining all needed information, the internal nozzle geometry is presented, the needle lift of the injector is characterized, high-speed Schlieren images of the injections are investigated, and the hydrogen mass flow and discharge coefficient through the nozzle are derived from high-speed pressure recordings. In the Methodology section, the experimental setup, the image processing and an empirical relationship to fit the jet penetration results are discussed. Subsequently, the results are presented and discussed, leading to a final conclusion in the last section of this paper.

\section{Methodology}
\subsection{Experimental Setup}
\begin{figure}
    \centering
    \includegraphics[width=0.7\linewidth]{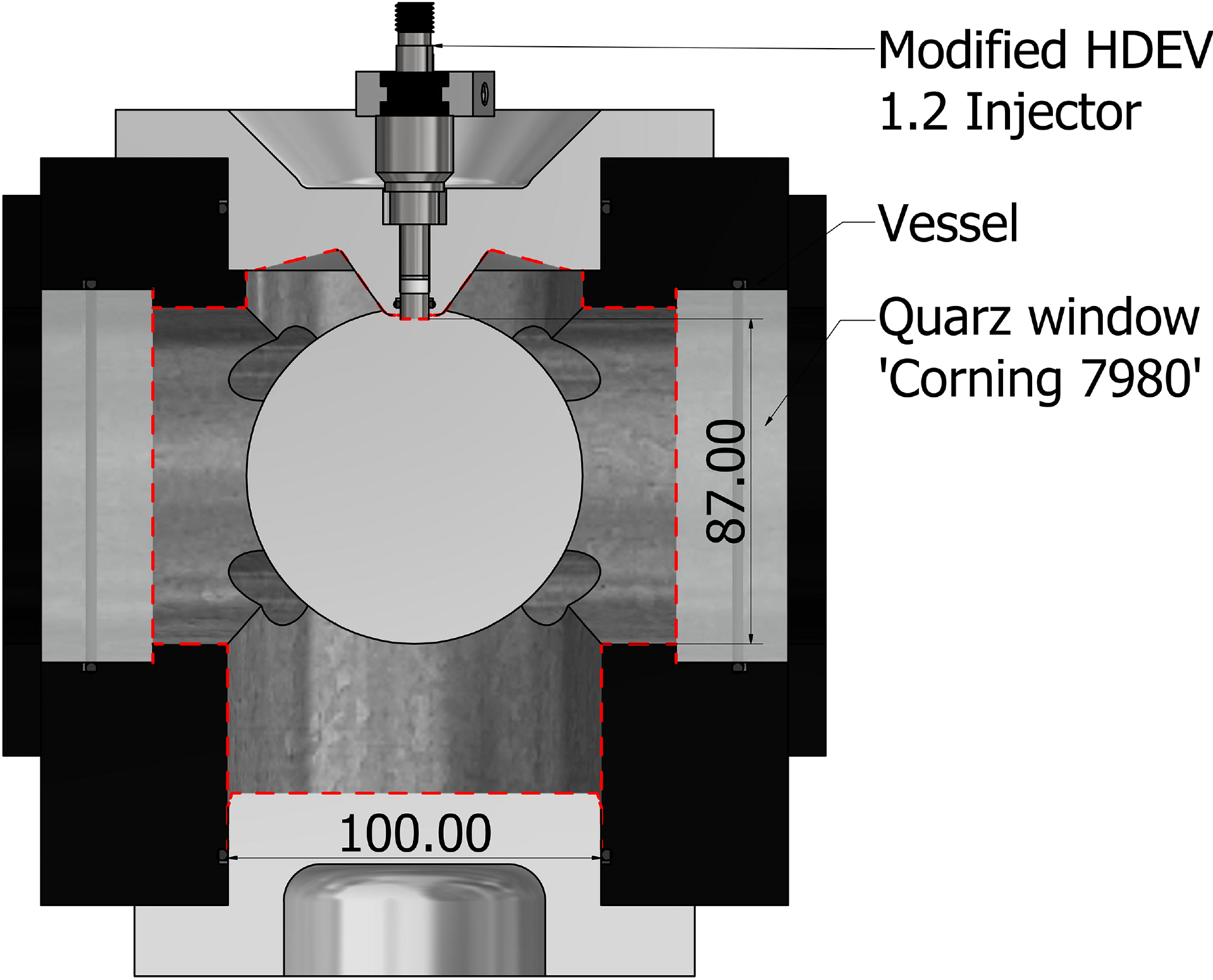}
    \caption{Cut-open 2D layout of the constant-volume setup and the top mounted injector. Red-dashed lines highlight the enclosed volume.}
    \label{fig:cell_CAD}
\end{figure}
The high-pressure hydrogen injection experiments are conducted in a non-heated constant-volume vessel, shown in \autoref{fig:cell_CAD}. This optically accessible vessel with a cylindrical chamber volume of $V_{\mathrm{c}} =$ $\SI{1800}{\cubic\cm}$ and 100-mm bore allows for precise control of the ambient gases at room temperature. The absolute ambient chamber pressure ($p_{\mathrm{a}}$) can be varied from 0.1-\SI{40}{bar}, allowing the comparison between inert argon and nitrogen environments. Hydrogen is injected through a modified Bosch HDEV 1.2 Gasoline Direct Injection (GDI) injector with an exit hole diameter ($d_\mathrm{n}$) of \SI{0.65}{\mm}, a maximum injection duration ($t_\mathrm{{inj}}$) of \SI{10}{\ms} and a maximum fuel pressure ($p_\mathrm{{f}}$) of \SI{100}{\bar}. Using the combination of a relative pressure transducer (Kistler 6041au20) and an absolute pressure transducer (Kistler 4005ba50), the pressure increase inside the vessel is measured from which the injected mass can be computed. In order to ease the control of injected mass, all measurement conditions in \autoref{table:expconditions} are chosen such that according to theory the flow is choked, i.e., $\mathrm{nPR} = \frac{p_\mathrm{f}}{p_\mathrm{a}}$ $> 1.899$.

\begin{table}
\caption{Range of experimental conditions, the conditions are chosen such that the choked flow limit applies ($\mathrm{nPR} > 1.899$).}
\label{table:expconditions}
\centering
\resizebox{0.5\linewidth}{!}{
\begin{tabular}{|ccc|}
\hline
\rowcolor[HTML]{343434} 
\multicolumn{3}{|c|}{\cellcolor[HTML]{343434}{\color[HTML]{FFFFFF} Experimental conditions}}                          \\
\rowcolor[HTML]{EFEFEF} 
{\color[HTML]{000000} nPR}              & {\color[HTML]{000000} -}   & {\color[HTML]{000000} 2 - 4 - 6 - 8 - 10}       \\
\rowcolor[HTML]{C0C0C0} 
{\color[HTML]{000000} $p_\mathrm{inj}$} & {\color[HTML]{000000} bar} & {\color[HTML]{000000} 20 - 40 - 60 - 80 - 100} \\
\rowcolor[HTML]{EFEFEF} 
{\color[HTML]{000000} $p_\mathrm{a}$}   & {\color[HTML]{000000} bar} & {\color[HTML]{000000} 2.0 - 40.0}              \\
\rowcolor[HTML]{C0C0C0} 
{\color[HTML]{000000} $t_\mathrm{inj}$} & {\color[HTML]{000000} ms}  & {\color[HTML]{000000} 2 - 6 - 10}              \\
\rowcolor[HTML]{EFEFEF} 
{\color[HTML]{000000} $T_\mathrm{f}$}   & {\color[HTML]{000000} K}   & {\color[HTML]{000000} 291}                     \\
\rowcolor[HTML]{C0C0C0} 
{\color[HTML]{000000} $T_\mathrm{a}$}   & {\color[HTML]{000000} K}   & {\color[HTML]{000000} 291}                     \\ \hline
\end{tabular}
}

\end{table}

\subsection{Injector geometry}

\begin{figure}
    \centering
    \begin{minipage}{0.30\textwidth}
              \centering

        \includegraphics[width=\linewidth]{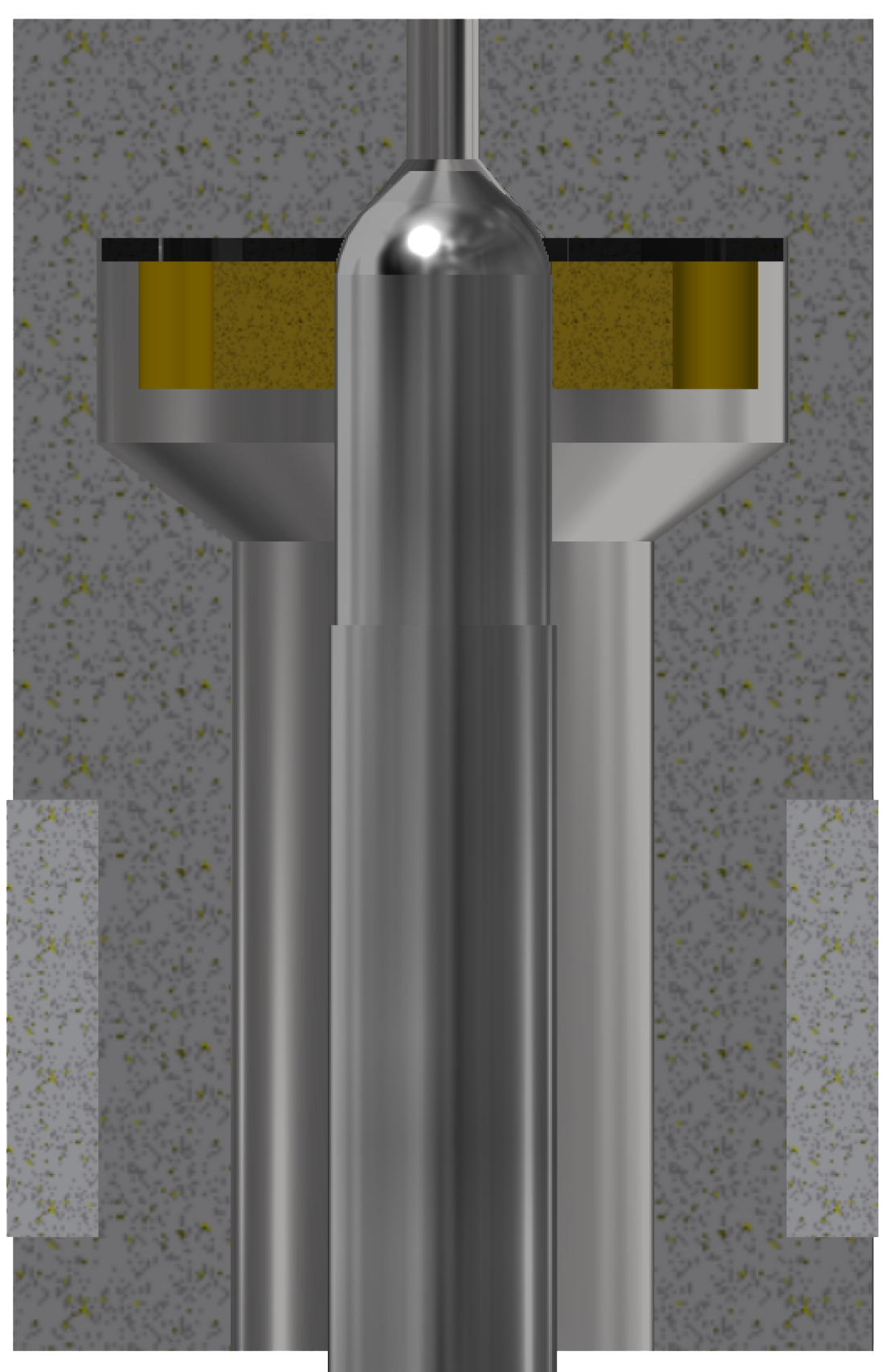}
    \end{minipage}%
    \begin{minipage}{0.4\textwidth}

        \includegraphics[width=.7\linewidth]{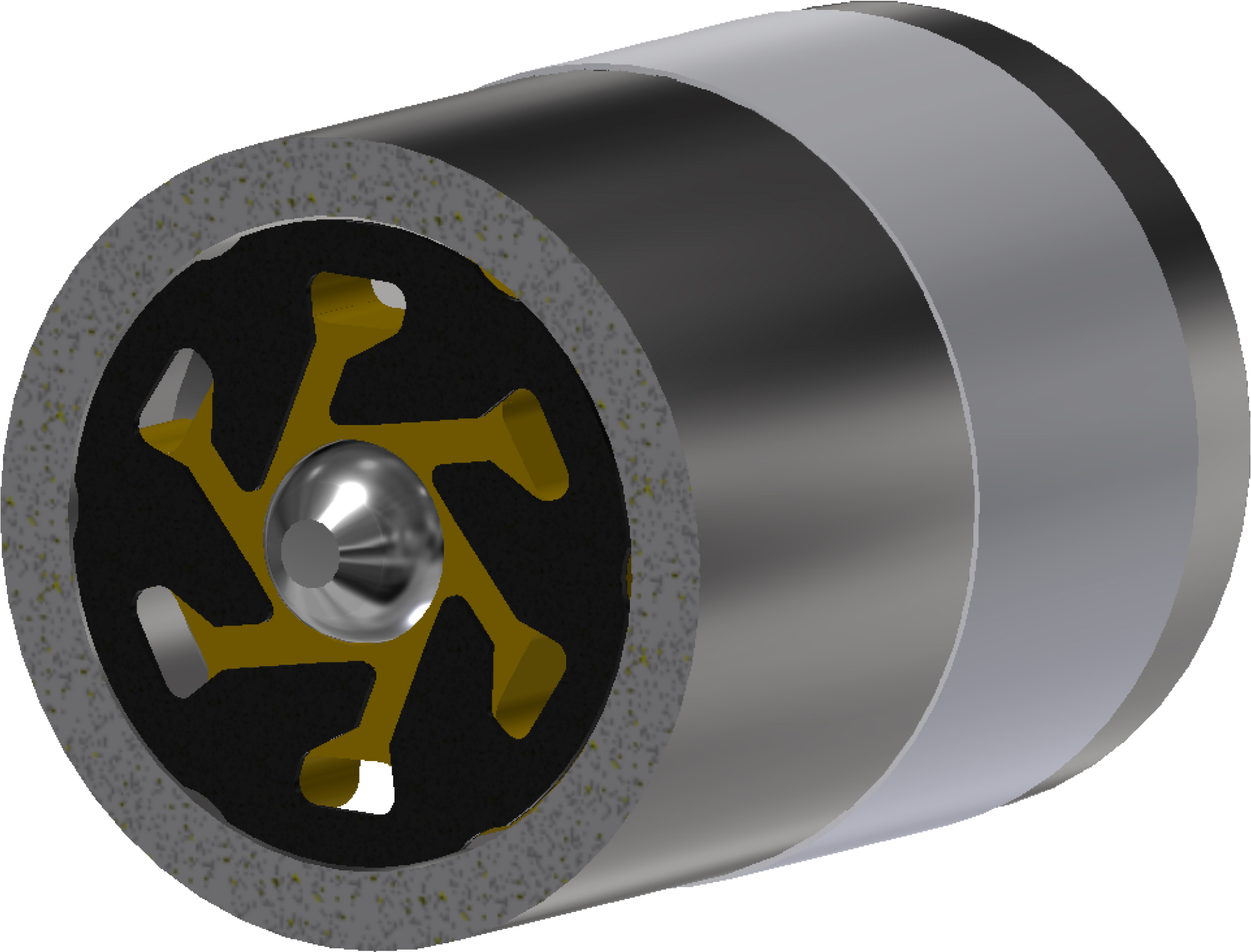}
        \centering

 \includegraphics[width=1\linewidth]{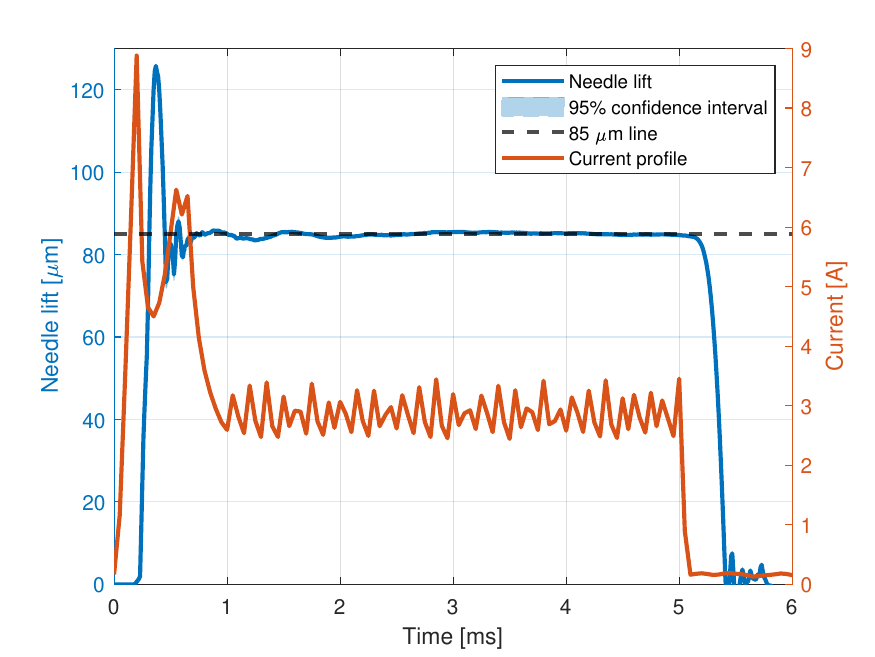}
        
    \end{minipage}
        \caption{Left panel: Cross-sectional view of the injector nozzle revealing the swirling geometry, needle and its conical seat. The right-top panel shows the pre-swirler (ochre) and swirling plate (black) inside the nozzle. The bottom panel shows the needle lift for ${t_\mathrm{inj} = \SI{5}{\ms}}$.}
        \label{fig:injectorcad}
\end{figure}

Using Computed Tomography (CT), the inner parts of the stainless steel injector nozzle are measured and presented in \autoref{fig:injectorcad}. The CT scan reveals a pre-swirler and a swirling plate before the conical seat at which the bullet-shaped needle rests. The HDEV 1.2 offers a threaded nozzle that is  modified to a straight single hole orifice, which is preferred to avoid line-of-sight interference of jets in an optically accessible constant-volume vessel. To measure the needle lift, a second HDEV injector was zinc sparked to reveal the needle during actuation with minimal intrusion of the needle seat. The needle lift is measured using a 200x zoom transmission microscope (Nexscope NE900, \SI{0.87}{\micro\m} per pixel) at \SI{100}{kHz} during a 5-ms electronic injection duration without gas flow at atmospheric pressure. \autoref{fig:injectorcad} shows an average needle lift of $\approx$ \SI{85}{\micro\m}, while overshoot and damping are visible during the opening event. After actuation, the needle bounces back from the seat during the closing event.

\subsection{Schlieren photography}
\begin{figure}
    \centering
    \includegraphics[width=\linewidth]{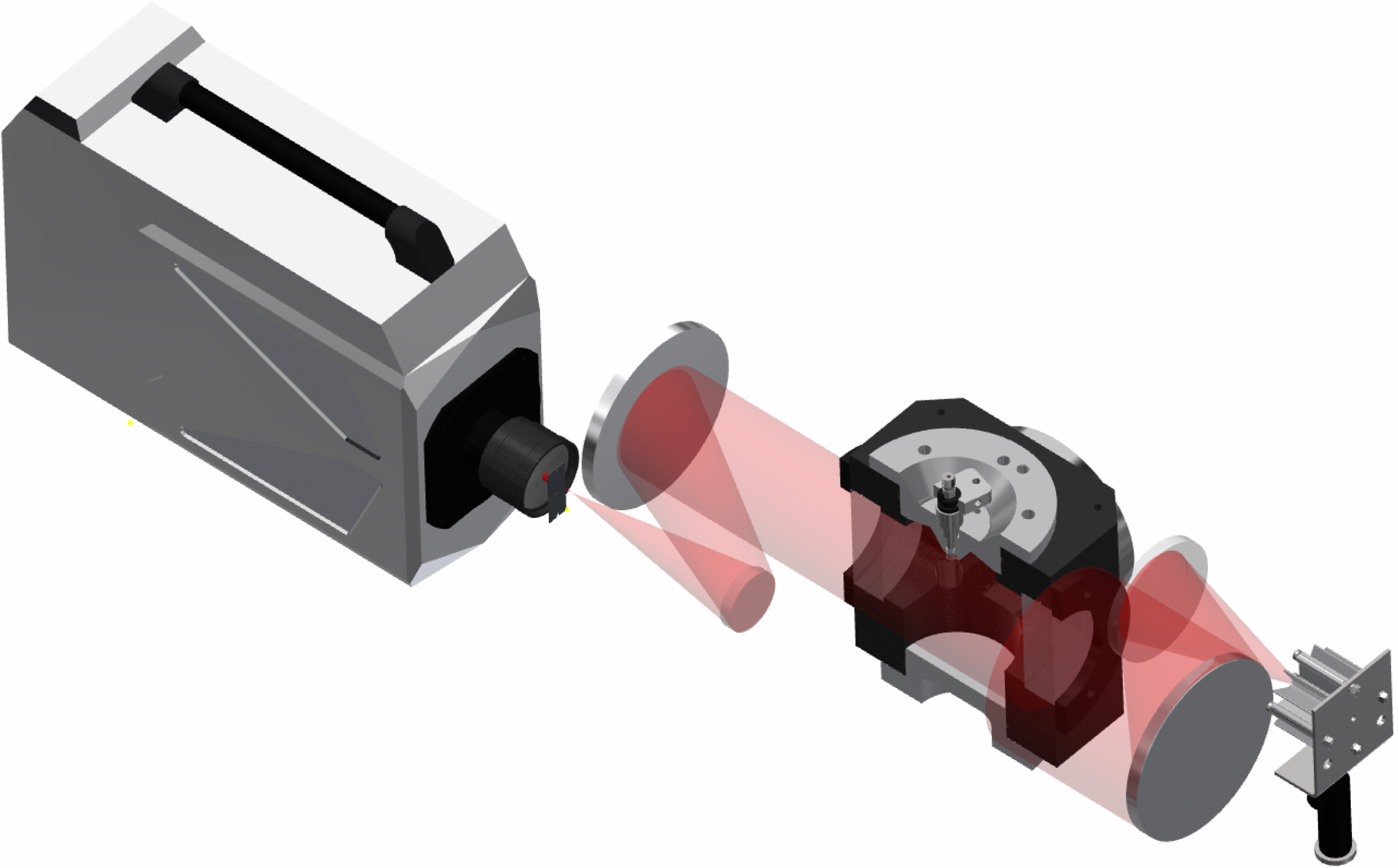}
    \caption{Cross section of the constant-volume vessel showcasing a Z-type Schlieren setup, from the CREE-LED to the Photron SA-Z camera. The (unperturbed) illumination path is shown in translucent red.}
    \label{fig:setup_CAD}
\end{figure}

Schlieren photography was used for high-speed imaging of the hydrogen jets, relying on the density gradients inside the vessel. Utilizing a Z-type setup as shown in \autoref{fig:setup_CAD}, light from a diaphragm masked 6W single CREE-LED (CW, \SI{680}{\nm}) is collimated by the first aluminum-coated parabolic mirror ($\mathrm{\frac{\lambda}{12}, f=\SI{1200}{\mm})}$ to direct the beam through the vessel and jet. The second parabolic mirror focuses the image back onto a straight vertical knife-edge cutting off rays that were diverted in the test section. The image is then captured by a Photron SA-Z through a Sigma 150-\SI{600}{mm} F5-6.3 lens for optimal use of the 1024x512 pixels ROI (\SI{0.085}{mm} per pixel) available at \SI{40}{\kHz}. The linear relationship between the refractive index $(n)$ and gas density $\mathrm{(\rho)}$ is expressed in terms of the Gladstone-Dale coefficient $(k)$ \cite{Settles2001SchlierenAS}:
\begin{equation}
    n-1 = k\rho 
\end{equation}

\begin{figure}
    \centering
    \includegraphics[width=\linewidth]{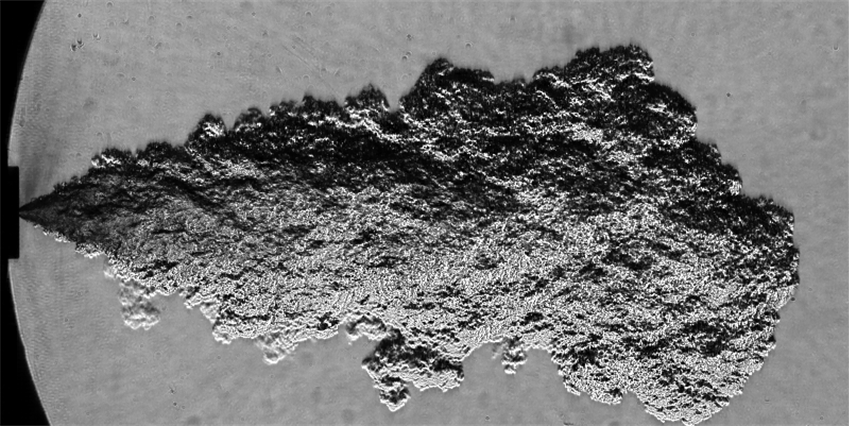}
    \caption{Still of the \SI{100}{bar} hydrogen jet which is injected into a \SI{40}{bar} argon ambient at \SI{5}{ms} aSOI. (In reality, the jet points downwards; the image is rotated 90 degrees.)}
    \label{fig:examplejet}
\end{figure}

Light rays change direction upon passing refractive index gradients. For small deflections, the angular ray deflection $\mathrm{(\epsilon_{x,y})}$ perpendicular to the unperturbed propogation direction are given by \cite{Settles2001SchlierenAS}: 
\begin{equation}
\epsilon_x = {\int_0}^L \frac{1}{n}  \frac{\partial n}{\partial x} \text{d}z, \epsilon_y = {\int_0}^L \frac{1}{n}  \frac{\partial n}{\partial y} \text{d}z,
\end{equation}
\noindent where $z$ is the 'normal' i.e., undisturbed ray propagation direction and $L$ is the distance from the source. Note that for a three dimensional jet with a different composition than the ambient, $L$ and $n$ are not necessarily constant along the $z$ direction, leading to line-of-sight quantification issues. Imaging hydrogen jets in an inert environment with a similar temperature requires a reasonable sensitivity of the system in order to sharply distinguish between the injected gas and the ambient, while too much sensitivity visualizes pressure waves surrounding the jet.

An enhanced and controlled sensitivity is achieved by the diaphragm that creates a small monochromatic light source which create the parallel light rays between the parabolic mirrors, while limiting chromatic abbreviations. The distance of the second mirror to the jet and the focusing distance determine the minimum sensitivity, while the amount of cut-out generated by the straight knife edge contributes to an increased amount of intensity difference in the horizontal direction. It was found that this combination resulted in the sharpest jet boundary with the best differentiation potential, as seen in \autoref{fig:examplejet}.

\subsection{Image processing}

\begin{figure}
    \centering
 \includegraphics[width=0.55\linewidth, angle =270 ]{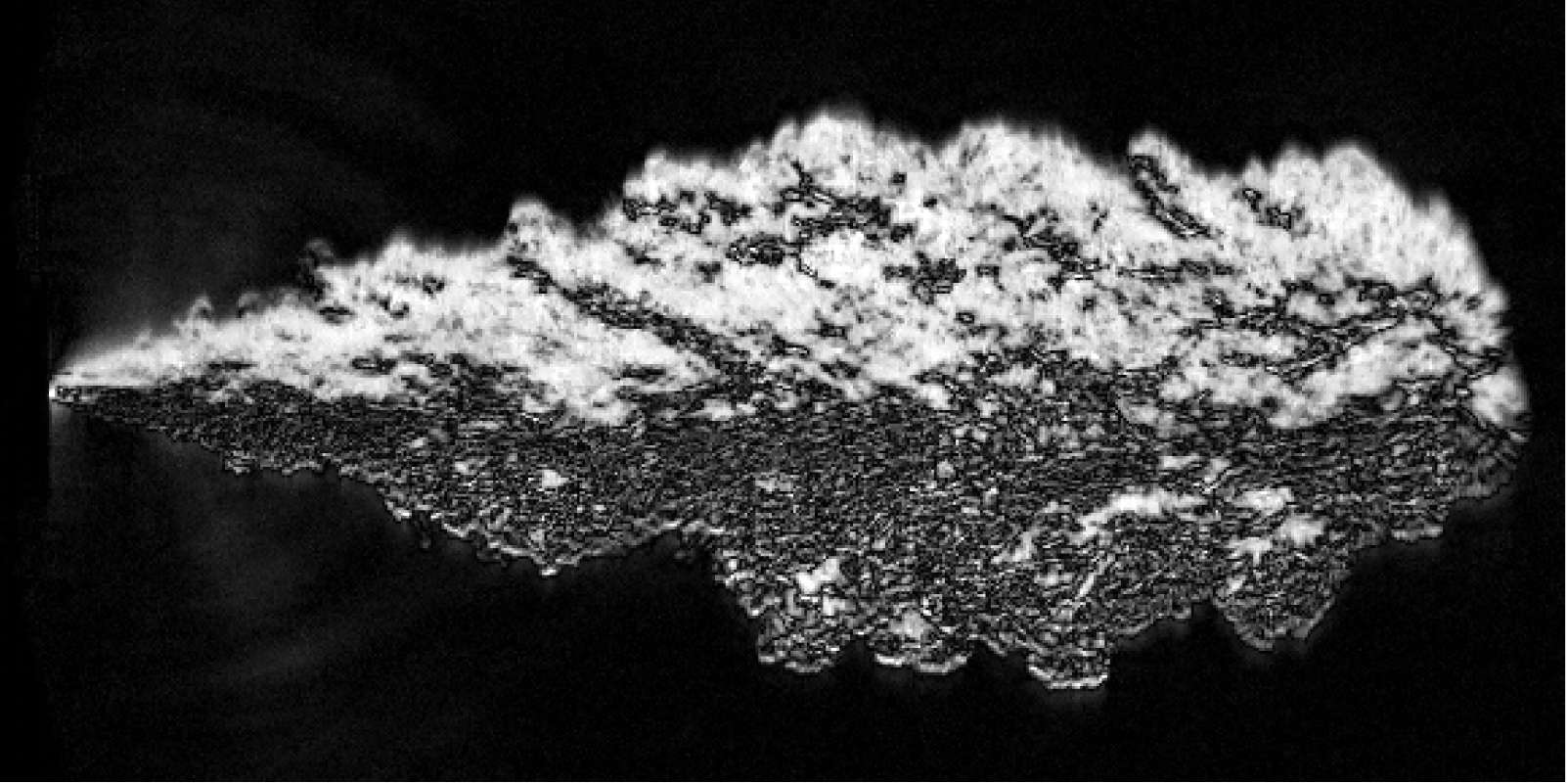}\hfill
 \includegraphics[width=0.55\linewidth, angle =270 ]{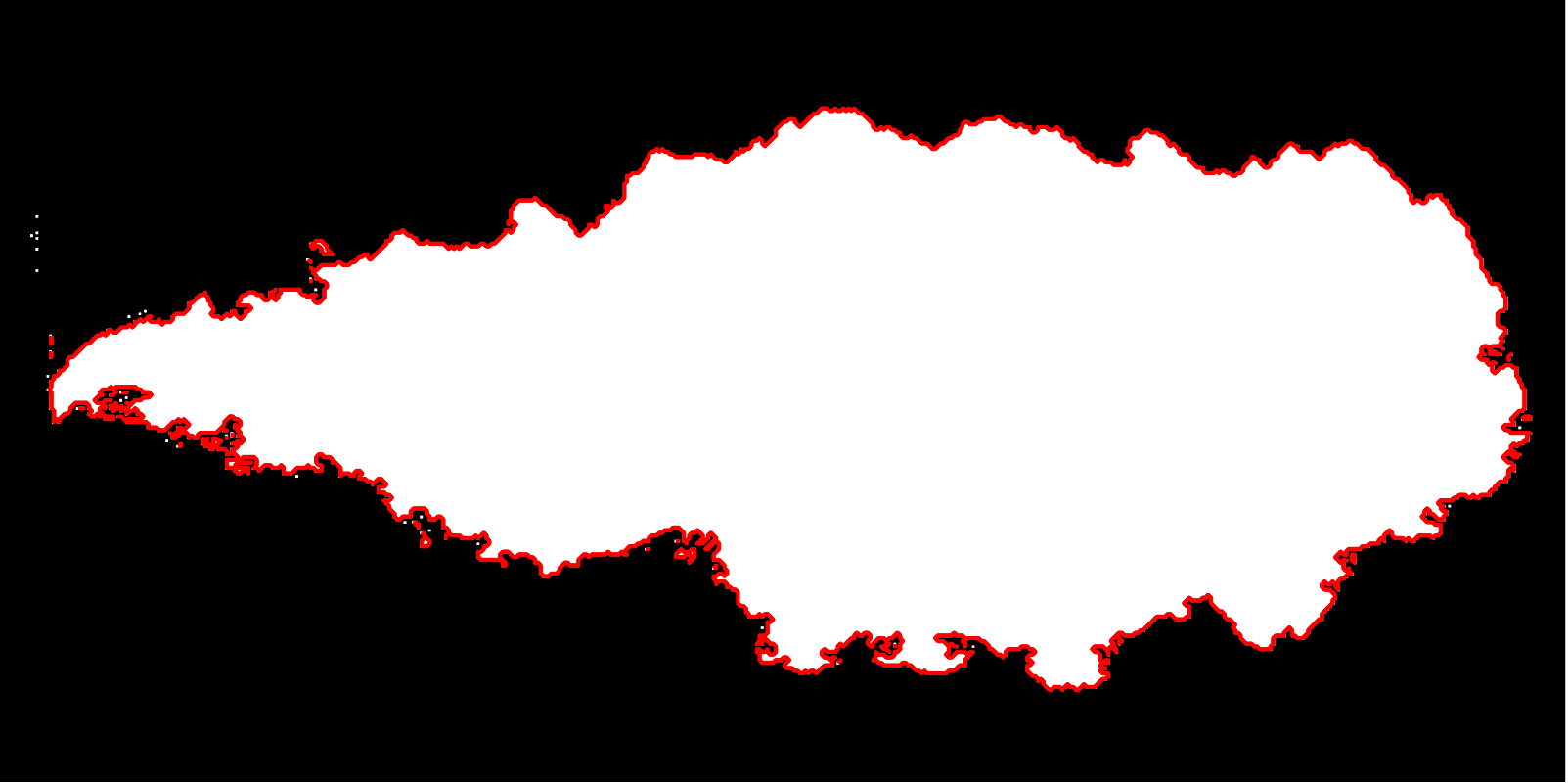}\hfill
 \includegraphics[width=0.55\linewidth, angle =270 ]{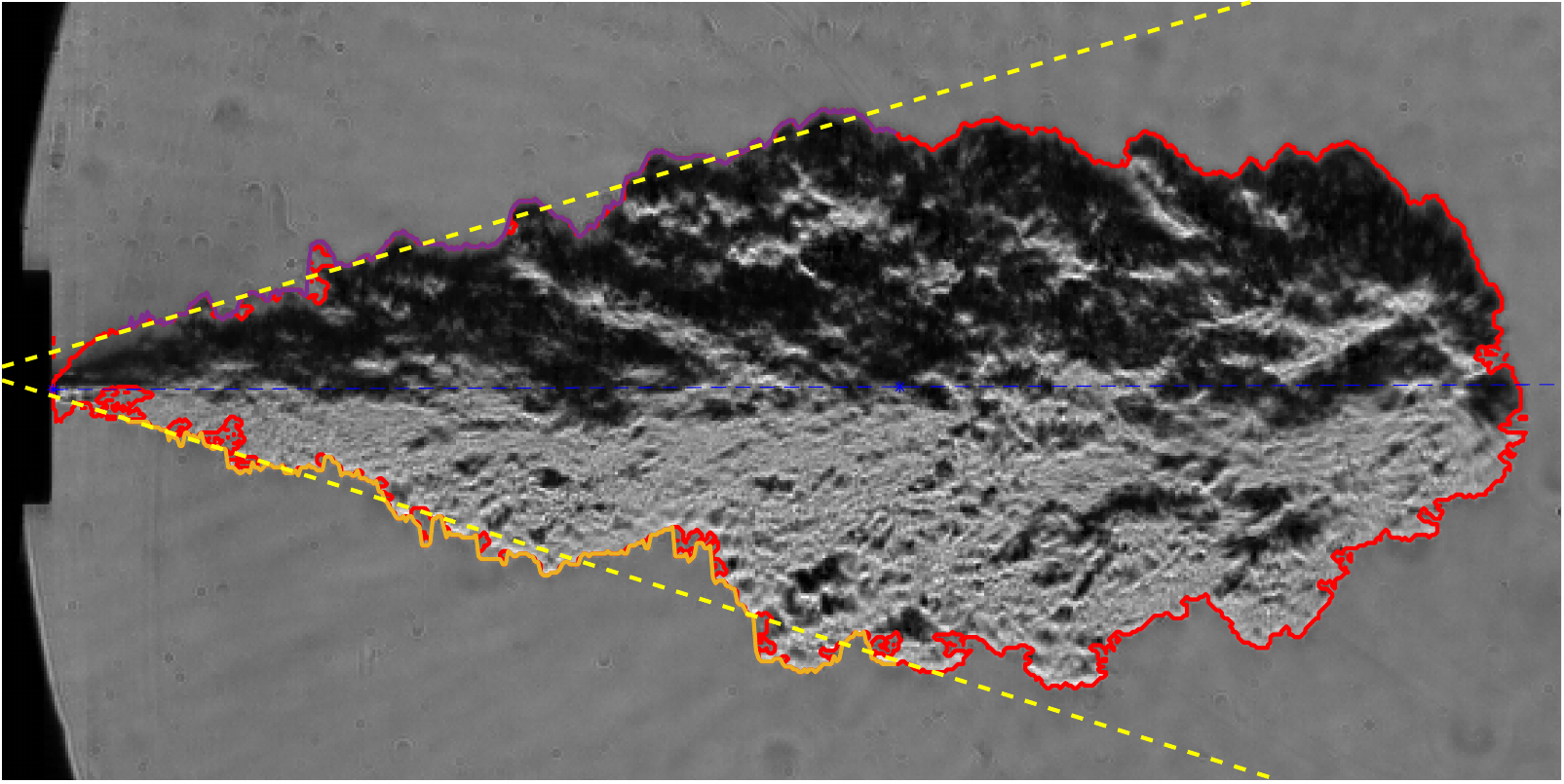}\hfill
    \caption{Left panel: absolute background subtracted Schlieren image, with visible pressure waves originating from the nozzle exit hole. The middle panel shows the binarized image, with the extracted red outline of the jet. Lastly, the original image shows in yellow and purple how a linear fit is made to calculate the overall jet angle. The images show a \SI{100}{bar} hydrogen jet into a \SI{10}{bar} argon filled ambient at \SI{1}{ms} aSOI.}
        \label{fig:postprocess_filtered}
\end{figure}

To determine jet penetration as a function of time, an image taken before the injection commences is subtracted from the jet images, a result is shown in the first panel of \autoref{fig:postprocess_filtered}. Subsequently, the absolute difference images are binarized using the MATLAB image processing toolbox using a fixed intensity threshold. This threshold is around 5-7 percent of the maximum intensity in the difference image to get a boundary detection of the jet periphery. A relatively low threshold is necessary given that density gradients caused by pressure waves are also visible in some conditions (\autoref{fig:postprocess_filtered}, first panel). Finally a mask is used in the post-processing routine to exclude pixels outside of the jet. The processed images are then analyzed further using a method described by Naber and Siebers \cite{naber}, which was mainly used for diesel sprays in their work. The images in this work are iteratively scanned from the injector tip until the jet head, i.e., where half of a projected curve (based on the computed jet angle) still overlaps with the ones in the binarized image. Due to the difference in geometry to conventional sprays, the gaseous jet tip is followed by using a quarter projected curve of the overlapping pixels. A slight modification was implemented for the post-processing of hydrogen jets, as the origin used for the jet angle calculation is different. In the case of a gas jet, the apparent jet origin lies within the nozzle. As a result, the jet angle will be significantly over-predicted ($>15 \mathrm{^{\circ}}$) if the origin is taken in the center of the nozzle exit. In addition to the significantly larger exit hole, the underexpanded jet also widens due to the aforementioned compressibility effects, which causes the apparent jet origin to shift upstream. 

With no robust method of measuring the exact width of the jet near the nozzle exit, because of the high-density gradients from pressure waves in this area, the jet angle is calculated by a linear fit on the jet outline starting two millimeter from the orifice exit all the way downstream to an axial location fixed as the weighted center point of the jet area. Despite the turbulent nature of the high-pressure jets, the fit follows the periphery of the jet closely, but starts to deviate after the weighted center point of the jet is reached. Fitting the jet periphery beyond this point would reduce the predicted angle significantly, as can be observed in the last panel of \autoref{fig:postprocess_filtered}. 

\subsection{Injected mass and discharge coefficient}
The injected mass ($m_\mathrm{inj}$) was computed by measuring the mean pressure rise in the vessel ($dp_\mathrm{a}$) after 100 consecutive injections including sufficient time for the gas inside the vessel to cool down to the temperature before the injection sequences ($T_\mathrm{a}$), and is averaged over three measurements. Using the ideal gas law the injected mass can be calculated as:

\begin{equation}
   m_{\mathrm{inj,H_2}} = dp_\mathrm{a} \cdot \frac{V_{\mathrm{c}}}{R_{\mathrm{H_2}}T_{\mathrm{a}}}
    \label{eq:massrise} 
\end{equation}

The theoretical idealized choked mass-flow ($\dot{m}_\mathrm{ideal}$) for a gas through a converging nozzle is based on assuming conservation of energy at stagnation conditions, frictionless flow, and ideal gas conditions to calculate the actual density ($\rho^*$) inside the nozzle exit hole, while assuming that the upstream velocity is choked at the speed of sound ($c$) \cite{turns_pauley_2020}. Combining these assumptions leads to:

\begin{equation}
\dot{m}_{ideal} = A_\mathrm{e}\cdot p_{\mathrm{f}} \sqrt{\frac{\gamma}{R T_\mathrm{f}} \left(\frac{2}{\gamma + 1} \right) ^\frac{\gamma+1}{\gamma-1}},
\label{eq:idealchokedmassflow}
    \end{equation}    
where $p_\mathrm{f}$ and $T_\mathrm{f}$ are the measured pressure and temperature in the fuel line, where the stagnation condition is assumed. The specific gas constant ($R_\mathrm{H_2}= \SI{4124}{J/(kg\cdot K)}$) and ratio of specific heat ($\mathrm{\gamma_{H_2} = 1.40}$) are taken for the upstream gas, at an assumed constant room temperature. Note that in the case of a choked flow, the mass flow through the nozzle exit plane does not depend on downstream conditions. Combining injection duration, injected mass, and idealized choked mass flow, the discharge coefficient ($C_d$) becomes: 

\begin{equation}
    C_\mathrm{d} = \frac{m_{\mathrm{inj}}}{\dot{m}_\mathrm{{ideal}} \cdot t_\mathrm{{inj}}}
    \label{eq:dischargecoeff_cd}
\end{equation}

\subsection{Fitting jet penetration}
\begin{figure}
    \centering
    \includegraphics[width=\linewidth]{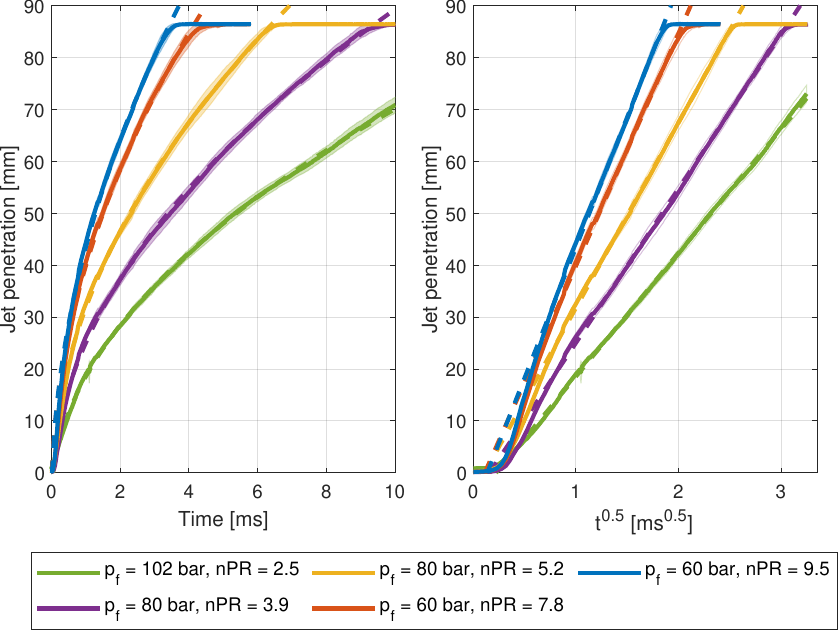}
    \caption{Examples of individual hydrogen jet penetrations in an argon ambient for various conditions. Solid lines are measurements, dashed lines are fits to \autoref{eq:jetpen_generic_formula}, plotted against time (left panel) and the square root of time (right panel). Measurements clip at 87 mm due to observation window size.}
    \label{fig:jepenmodel_generic}
\end{figure}

An effort was made to obtain an empirical correlation for the jet penetration ($S$) results. In line with continuous jet momentum theory the gaseous jet is assumed to have a constant rate of entrainment \cite{ricou_spalding_1961}, which leads to $S \propto t^{1/2}$, resulting in a general equation in the form of:
\begin{equation}
S(t) = A\cdot t^{1/2} + B
    \label{eq:jetpen_generic_formula}
\end{equation}

In order to fit a relationship to the measurement data \autoref{eq:jetpen_generic_formula} is used, where $t = 0$ is the time where $S = \SI{1.5}{mm}$, while ${A}$ and ${B}$ are fitting parameters introduced as overall gain ($A$) and offset ($B$) for the opening needle and start of flow. \autoref{fig:jepenmodel_generic} shows that jet penetration is fitted very closely from \SI{20}{mm} onwards. It should be noted that there is no further physical background for \autoref{eq:jetpen_generic_formula}, but the fitting parameters are purely based on good agreement with the obtained measurement data.

\section{Results and Discussion}
\subsection{Jet penetration}
\begin{figure}
    \centering
\includegraphics[width=\linewidth]{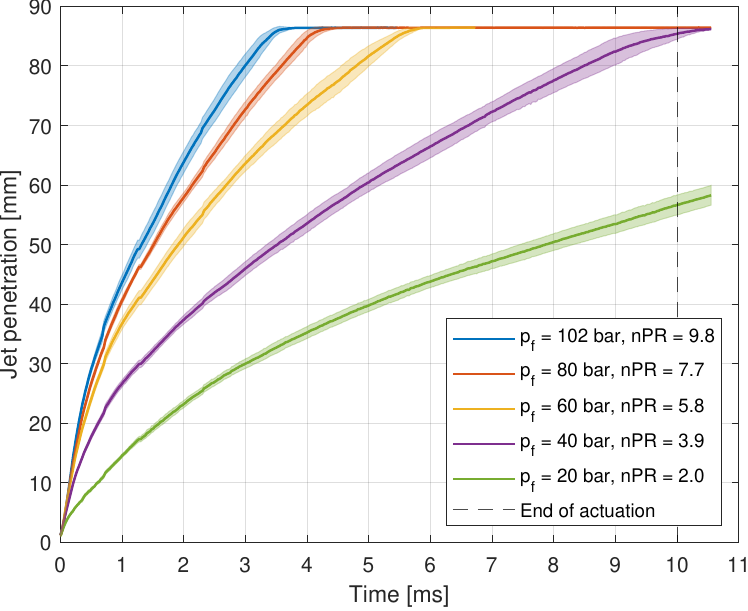}
    \caption{Jet penetration and 95\% confidence intervals (shaded, based on 10 repetitions) of hydrogen injections at different injection pressures into an argon ambient with $p_\mathrm{a} = \SI{10}{bar}.$}
    \label{fig:jetpen_10bar}
\end{figure}

Jet penetration is measured and averaged over ten measurements at the same conditions. The ensemble average of jet penetration and 95\% confidence interval are presented in \autoref{fig:jetpen_10bar} for different injection pressures into 10 bar of argon. Clearly, jet penetration is faster for higher nPR in accordance with the theory of underexpanded jets discussed in Franquet et al.\ \cite{FRANQUET201525}. In their work, they show particularly how the near-field geometry of the jet is determined by the pressure ratio, while Tsujimura et al.\ and Ouellette provided empirical relationships that show how the far-field of the jet is dominated by the upstream conditions and the chamber density ($\rho_\mathrm{a}$) \cite{tsujimura,Ouellette_1996}. As ambient temperature and composition are essentially constant, the chamber density depends solely on $p_\mathrm{a}$. 

\begin{figure}
    \centering
\includegraphics[width=\linewidth]{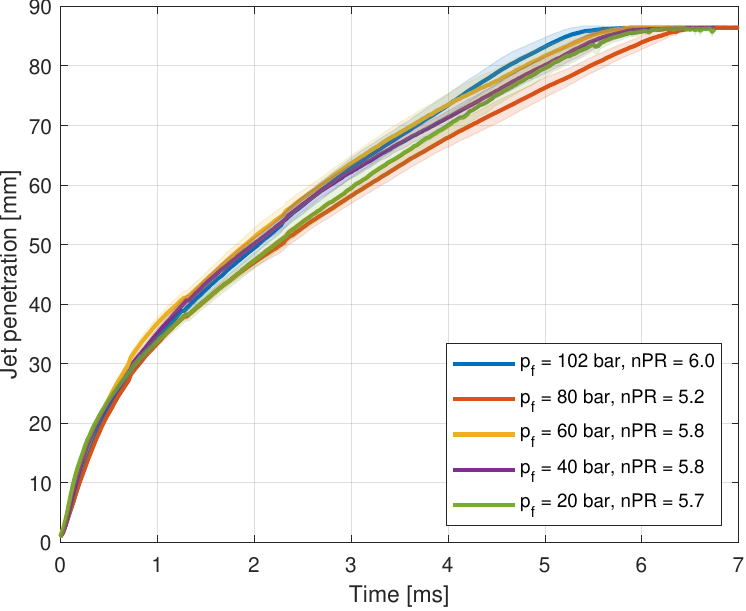}
    \caption{Jet penetration and 95\% confidence intervals with a pressure ratio between 5 and 6 in an argon ambient. }
    \label{fig:jetpen_equalnpr}
\end{figure}

\autoref{fig:jetpen_equalnpr} shows jet penetration for different fuel and chamber pressures, but with similar nPR. Slight variations in the pressure ratio correlate with small variations in jet penetration. When taking the $p_\mathrm{f} = \SI{80.3}{bar}$ (red line) measurement as an example, it clearly has the slowest jet penetration due to the lower pressure ratio, while it does not have the lowest injection pressure. This trend holds for all $\mathrm{nPR>1.899}$ measurements studied in this setup.  

\subsection{Jet angle}

\begin{figure}
    \centering
    \includegraphics[width=\linewidth]{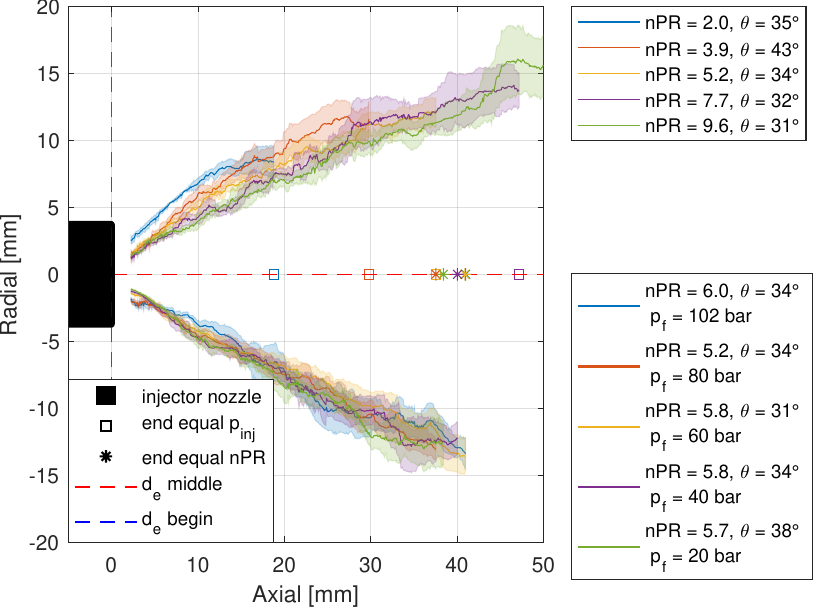}
    \caption{Jet boundaries (solid lines) and confidence intervals (shading) for various conditions at 1ms aSOI. The top half shows conditions with $p_\mathrm{inj} = \SI{80}{bar}$ at varying pressure ratios, while the lower half shows equal pressure ratio conditions with varying fuel pressure. The symbols at the centerline indicate the location of the weighted centre point of the jet, where the fit ends.}
    \label{fig:jetangletwice}
\end{figure}

The continuous jet momentum models from Naber and Siebers\ \cite{naber} and Tsujimura et al.\ \cite{tsujimura} show that the angle of the jet ($\theta$) directly influences jet penetration , due to the geometric assumption of a conical, axisymmetric jet. Efforts were made to quantitatively measure and compare the angle by taking the average of the jet boundary contours of ten measurements for each condition. The result in \autoref{fig:jetangletwice} shows one half of these jet boundary contours for both the varying nPR (top) and equal nPR conditions (bottom). The confidence intervals in \autoref{fig:jetangletwice} are relatively large for these highly turbulent jets and they increase with the distance from the nozzle. 

At varying nPR (top half) there are some slight deviations in the jet angle visible in the 5-20 mm region, where the jet angle becomes smaller for increasing nPR. This decrease in jet angle is accompanied by a faster penetration, as denoted by the squares on the (red) axis. The mean fitted angles vary between 30 and 43 degrees, showing significant differences in jet angle. The $\mathrm{nPR = 2}$ condition shows relatively slow penetration and a wider jet near the nozzle, which could be due to the low pressure ratio, where the actual pressure inside the nozzle is lower than the choked flow limit. When the pressure ratio is kept constant (bottom half of \autoref{fig:jetangletwice}) the 95\% confidence intervals from the jet boundary are overlapping. The mean jet angle was fitted, but the mean values are in the range of 31 - 38 degrees without a clear pattern. The jet boundary seems to be primarily governed by nPR and its influence is further discussed in \autoref{sub:emp}.

\subsection{Injected mass \& discharge coefficient}
\begin{table}
\caption{Discharge coefficient and injected mass for different conditions of the hydrogen injections into an argon ambient at room temperature ($T_\mathrm{f}=T_\mathrm{a}= \SI{291}{K}$).}
\label{table:massofinjection}
\centering
\resizebox{0.5\linewidth}{!}{

\begin{tabular}{|cccccc|}
\hline
\rowcolor[HTML]{343434} 
\multicolumn{4}{|c}{\cellcolor[HTML]{343434}{\color[HTML]{FFFFFF} Measured conditions}}                                                                                                                                                                                                                                                                                      & \multicolumn{2}{c|}{\cellcolor[HTML]{343434}{\color[HTML]{FFFFFF} Results}}                                                                                                          \\ \hline
\rowcolor[HTML]{656565} 
{\color[HTML]{FFFFFF} \begin{tabular}[c]{@{}c@{}}$p_\mathrm{f}$\\ {[}bar{]}\end{tabular}} & {\color[HTML]{FFFFFF} \begin{tabular}[c]{@{}c@{}}$p_\mathrm{a}$\\ {[}bar{]}\end{tabular}} & {\color[HTML]{FFFFFF} \begin{tabular}[c]{@{}c@{}}$t_\mathrm{inj}$\\ {[}ms{]}\end{tabular}} & {\color[HTML]{FFFFFF} \begin{tabular}[c]{@{}c@{}}$n_\mathrm{a}$\\ {[}-{]}\end{tabular}} & {\color[HTML]{FFFFFF} \begin{tabular}[c]{@{}c@{}}$m_\mathrm{inj}$\\ {[}mg{]}\end{tabular}} & {\color[HTML]{FFFFFF} \begin{tabular}[c]{@{}c@{}}$C_\mathrm{d}$\\ {[}-{]}\end{tabular}} \\ \hline
\rowcolor[HTML]{C0C0C0} 
100.6                                                                                     & 20.2                                                                                      & 10                                                                                         & Ar                                                                                      & 9.18                                                                                       & 0.44                                                                                    \\
\rowcolor[HTML]{EFEFEF} 
100.6                                                                                     & 20.1                                                                                      & 6                                                                                          & Ar                                                                                      & 5.49                                                                                       & 0.44                                                                                    \\
\rowcolor[HTML]{C0C0C0} 
100.6                                                                                     & 20.0                                                                                      & 2                                                                                          & Ar                                                                                      & 1.74                                                                                       & 0.42                                                                                    \\
\rowcolor[HTML]{EFEFEF} 
80.2                                                                                      & 10.1                                                                                      & 10                                                                                         & Ar                                                                                      & 7.24                                                                                       & 0.43                                                                                    \\
\rowcolor[HTML]{C0C0C0} 
42.6                                                                                      & 10.1                                                                                      & 10                                                                                         & Ar                                                                                      & 3.71                                                                                       & 0.42                                                                                    \\ \hline

\end{tabular}}

\end{table}

\autoref{table:massofinjection} gives an overview of the injected mass and discharge coefficient. The results illustrate that the mass flow through the injector is depending solely on the upstream gas conditions, as varying chamber pressure only leads to variations within one percent of the discharge coefficient. This is in agreement with \autoref{eq:idealchokedmassflow}, which predicts that $\dot{m}_\mathrm{ideal}$ is independent of chamber conditions, showing that the experimental conditions indeed correspond to choked flow in the nozzle. Furthermore, the results in the table show almost identical discharge coefficients, meaning that the experimental $m_\mathrm{inj}$ follows a linear trend depending on both fuel injection duration and fuel pressure. 

While the injected mass at the highest allowable $p_\mathrm{f}$ for this injector is small compared to conventional diesel injections, this could be solved by utilizing multiple or larger holes to increase the flow area and improving the flow efficiency ($C_\mathrm{d}$) through the nozzle by optimizing the geometry (i.e., increasing needle lift enlarging/removing the swirl path, and removing sharp edges).

\subsection{Argon vs. nitrogen ambient}

\begin{table}
\caption{Comparison for $\mathrm{C_d}$ and $\mathrm{m_{inj}}$ into argon and nitrogen for different injection timings at room temperature ($T_\mathrm{f}=T_\mathrm{a}= \SI{291}{K}$).}
\label{table:massofinjectionArvsN2}
\centering
\resizebox{0.5\linewidth}{!}{

\begin{tabular}{cccccc}
\hline
\rowcolor[HTML]{343434} 
\multicolumn{4}{c}{\cellcolor[HTML]{343434}{\color[HTML]{FFFFFF} Measured conditions}}                                                                                                                                                                                                                                                                                                                                    & \multicolumn{2}{c}{\cellcolor[HTML]{343434}{\color[HTML]{FFFFFF} Results}}                                                                                                                                                        \\ \hline
\rowcolor[HTML]{656565} 
\multicolumn{1}{|c}{\cellcolor[HTML]{656565}{\color[HTML]{FFFFFF} \begin{tabular}[c]{@{}c@{}}$p_\mathrm{f}$\\ {[}bar{]}\end{tabular}}} & {\color[HTML]{FFFFFF} \begin{tabular}[c]{@{}c@{}}$p_\mathrm{a}$\\ {[}bar{]}\end{tabular}} & {\color[HTML]{FFFFFF} \begin{tabular}[c]{@{}c@{}}$t_\mathrm{inj}$\\ {[}ms{]}\end{tabular}} & {\color[HTML]{FFFFFF} \begin{tabular}[c]{@{}c@{}}$n_\mathrm{a}$\\ {[}-{]}\end{tabular}} & {\color[HTML]{FFFFFF} \begin{tabular}[c]{@{}c@{}}$m_\mathrm{inj}$\\ {[}mg{]}\end{tabular}} & \multicolumn{1}{c|}{\cellcolor[HTML]{656565}{\color[HTML]{FFFFFF} \begin{tabular}[c]{@{}c@{}}$C_\mathrm{d}$\\ {[}-{]}\end{tabular}}} \\
\rowcolor[HTML]{C0C0C0} 
\multicolumn{1}{|c}{\cellcolor[HTML]{C0C0C0}{\color[HTML]{000000} 99.7}}                                                               & {\color[HTML]{000000} 10.2}                                                               & {\color[HTML]{000000} 10}                                                                  & {\color[HTML]{000000} $\mathrm{N_2}$}                                                   & {\color[HTML]{000000} 8.97}                                                                & \multicolumn{1}{c|}{\cellcolor[HTML]{C0C0C0}{\color[HTML]{000000} 0.43}}                                                             \\
\rowcolor[HTML]{EFEFEF} 
\multicolumn{1}{|c}{\cellcolor[HTML]{EFEFEF}{\color[HTML]{000000} 100.6}}                                                              & {\color[HTML]{000000} 10.2}                                                               & {\color[HTML]{000000} 10}                                                                  & {\color[HTML]{000000} Ar}                                                               & {\color[HTML]{000000} 9.04}                                                                & \multicolumn{1}{c|}{\cellcolor[HTML]{EFEFEF}{\color[HTML]{000000} 0.43}}                                                             \\
\rowcolor[HTML]{C0C0C0} 
\multicolumn{1}{|c}{\cellcolor[HTML]{C0C0C0}{\color[HTML]{000000} 99.7}}                                                               & {\color[HTML]{000000} 10.2}                                                               & {\color[HTML]{000000} 6}                                                                   & \cellcolor[HTML]{C0C0C0}{\color[HTML]{000000} $\mathrm{N_2}$}                           & {\color[HTML]{000000} 5.33}                                                                & \multicolumn{1}{c|}{\cellcolor[HTML]{C0C0C0}{\color[HTML]{000000} 0.43}}                                                             \\
\rowcolor[HTML]{EFEFEF} 
\multicolumn{1}{|c}{\cellcolor[HTML]{EFEFEF}{\color[HTML]{000000} 100.6}}                                                              & {\color[HTML]{000000} 10.2}                                                               & {\color[HTML]{000000} 6}                                                                   & {\color[HTML]{000000} Ar}                                                               & {\color[HTML]{000000} 5.39}                                                                & \multicolumn{1}{c|}{\cellcolor[HTML]{EFEFEF}{\color[HTML]{000000} 0.42}}                                                             \\
\rowcolor[HTML]{C0C0C0} 
\multicolumn{1}{|c}{\cellcolor[HTML]{C0C0C0}{\color[HTML]{000000} 99.7}}                                                               & {\color[HTML]{000000} 9.96}                                                               & {\color[HTML]{000000} 2}                                                                   & \cellcolor[HTML]{C0C0C0}{\color[HTML]{000000} $\mathrm{N_2}$}                           & {\color[HTML]{000000} 1.69}                                                                & \multicolumn{1}{c|}{\cellcolor[HTML]{C0C0C0}{\color[HTML]{000000} 0.41}}                                                             \\ \hline

\end{tabular}}

\end{table}

\begin{figure}
    \centering
    \includegraphics[width=\linewidth]{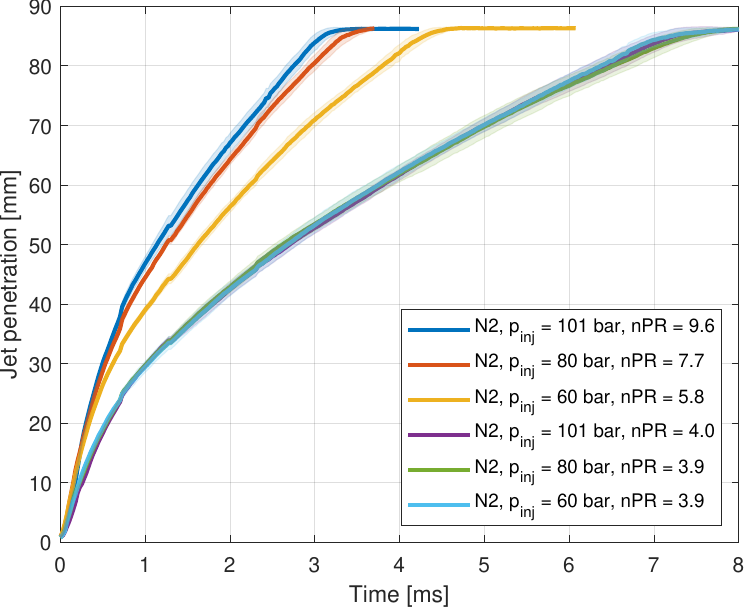}
    \caption{Jet penetration results for nitrogen environments at varying conditions.}
    \label{fig:N2_jetpenresults}   
\end{figure}

To simulate differences between argon and air as working fluids, nitrogen was used to create a pressurized inert ambient. Assuming the ideal mass flow conditions, there should be no difference in injected mass when changing the composition of the ambient gas ($n_\mathrm{a}$). This is indeed borne out by the experimental result, summarized in \autoref{table:massofinjectionArvsN2}.

\begin{figure}
    \centering
\includegraphics[width=\linewidth]{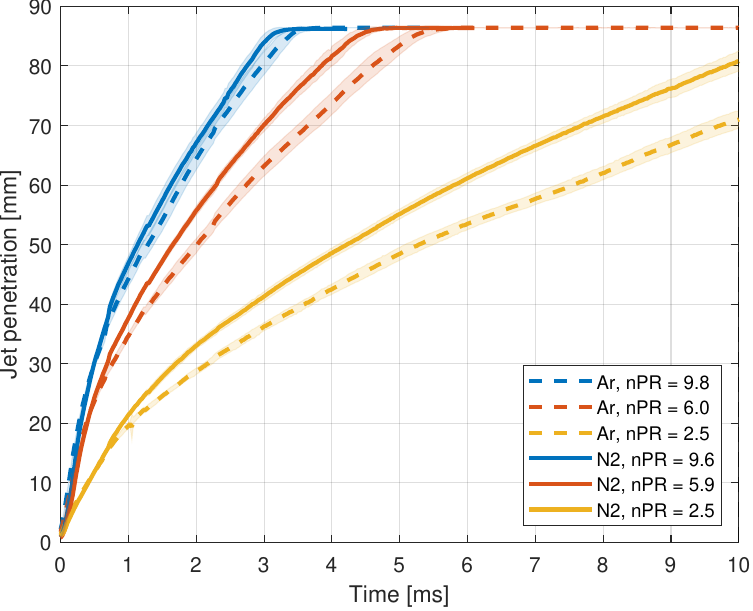}
    \caption{Jet penetration results in argon (dashed lines) and nitrogen (solid lines) at different ambient pressures with $p_\mathrm{f} = \SI{100}{bar}$.}
    \label{fig:ArgvsN2_100bar}
\end{figure}

\begin{figure}
\centering
\begin{minipage}{.48\textwidth}
    \centering
    \includegraphics[height=0.207\textheight]{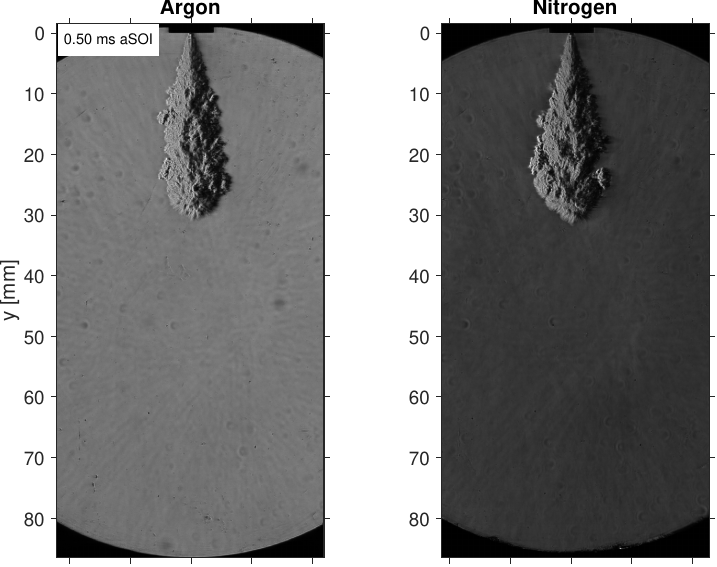}
\end{minipage}
\hfill
\begin{minipage}{.48\textwidth}
    \centering
     \includegraphics[height=0.207\textheight]{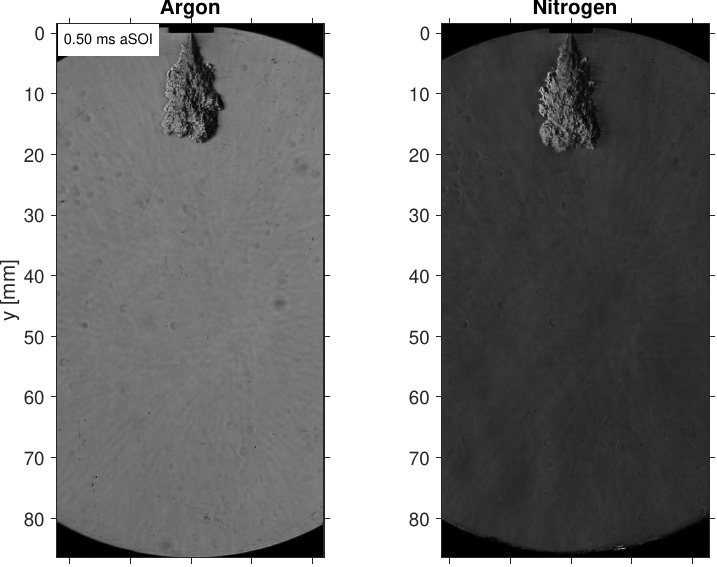}
\end{minipage}

\begin{minipage}{.48\textwidth}
    \centering
    \includegraphics[height=0.2\textheight]{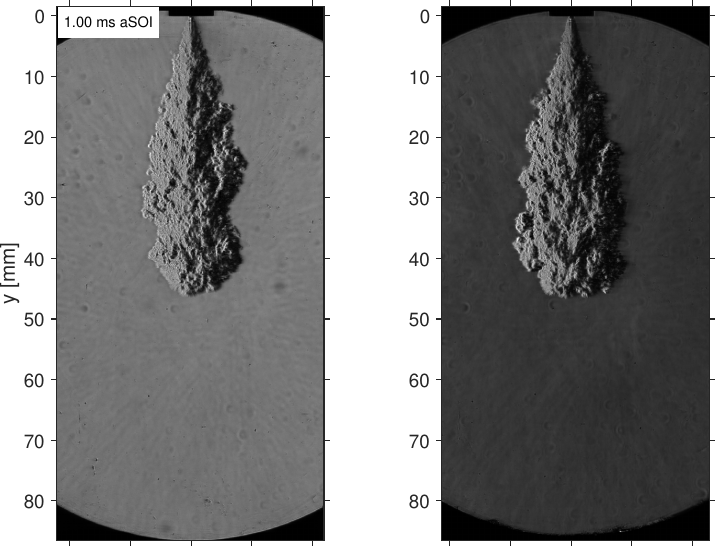}
\end{minipage}
\hfill
\begin{minipage}{.48\textwidth}
    \centering
     \includegraphics[height=0.2\textheight]{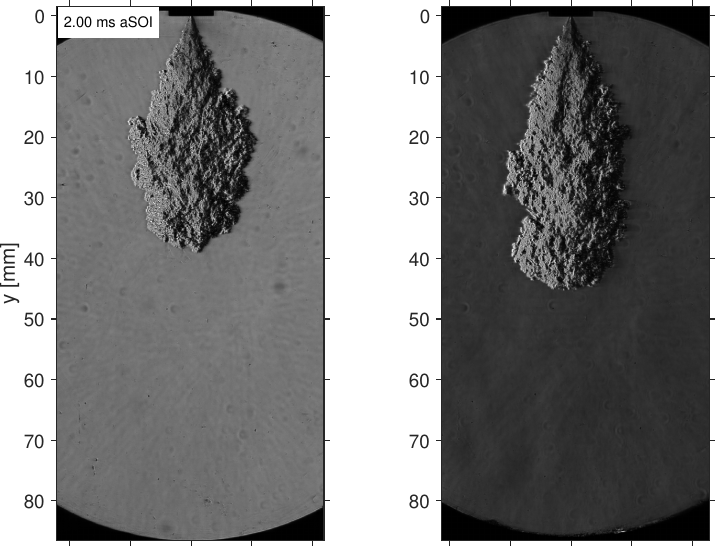}
\end{minipage}

 \begin{minipage}{.48\textwidth}
    \centering
    \includegraphics[height=0.2\textheight]{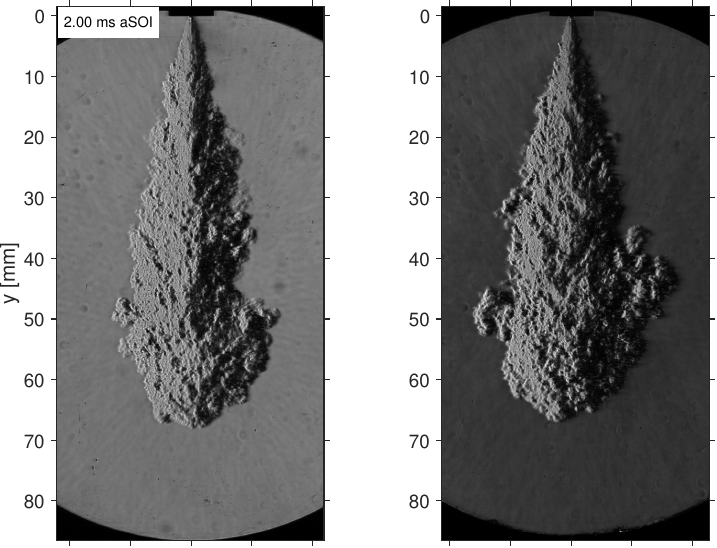}
\end{minipage}
\hfill
\begin{minipage}{.48\textwidth}
    \centering
     \includegraphics[height=0.2\textheight]{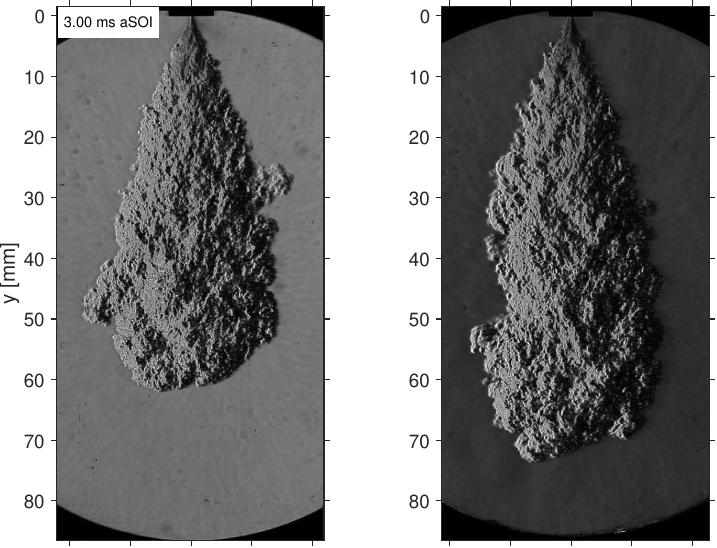}
\end{minipage}

 \begin{minipage}[t]{.48\textwidth}
    \centering
     \includegraphics[width=0.87\textwidth]{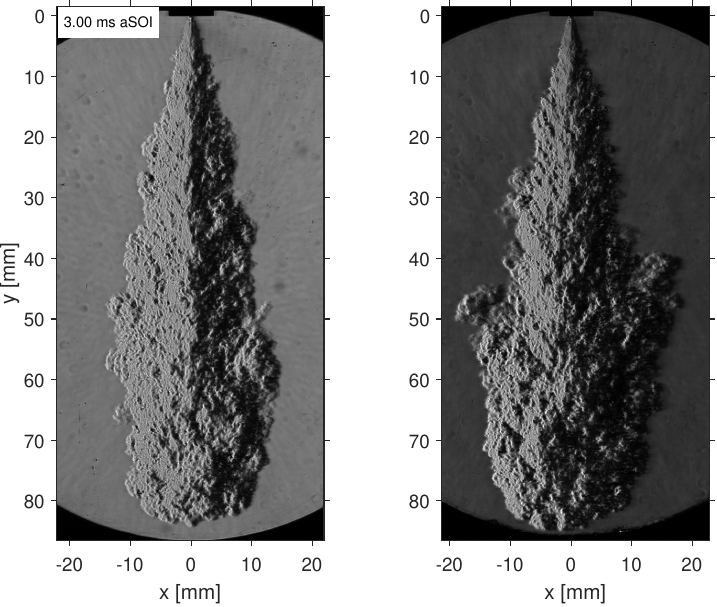}

    \caption{Schlieren images for a 100-bar $\mathrm{H_{2}}$ injection into \SI{10}{bar} of argon (left panels) or nitrogen (right panels). Note that minor deviations in diaphragm and knife edge positioning yield slight variations in light intensity of the background, but no impact on the jet boundary detection is expected.}
    \label{fig:qualschlieren100_10}
\end{minipage}
\hfill
\begin{minipage}[t]{.48\textwidth}
   \centering
    \includegraphics[width=0.87\textwidth]{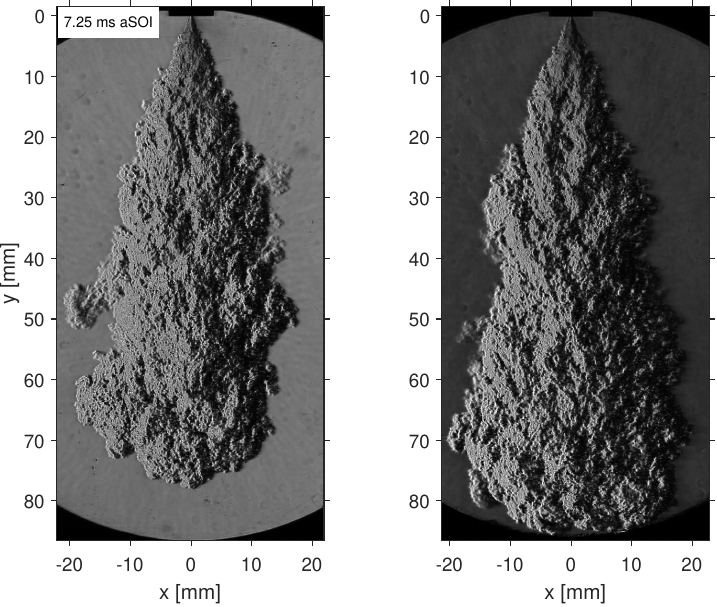}
      \caption{Schlieren images for an 80-bar $\mathrm{H_{2}}$ injection into \SI{20}{bar} of argon (left panels) or nitrogen (right panels).}
    \label{fig:qualschlieren80_20}
\end{minipage}    
\end{figure}

When evaluating jet penetration results for the nitrogen ambient in \autoref{fig:N2_jetpenresults}, the same trend as with argon is evident: jet penetration is mainly affected by nPR. Increasing nPR implies a faster penetration and equal nPR yields a similar jet penetration, while $m_{\mathrm{inj}}$ depends only on $p_\mathrm{f}$. However, \autoref{fig:ArgvsN2_100bar} shows that mean jet penetration is faster for nitrogen (solid lines) than for the argon (dashed lines) ambient, which holds for all measured conditions. Time sequenced Schlieren images of hydrogen injections into argon and nitrogen are shown in \autoref{fig:qualschlieren100_10}. At these conditions the jets behave very similarly with overlapping 95\% confidence intervals in \autoref{fig:ArgvsN2_100bar}. When evaluating $\mathrm{nPR = 4}$ in \autoref{fig:qualschlieren80_20}, the difference in mean jet penetration is much greater than the variation, showing an always faster jet in nitrogen at lower nPR.

\begin{figure}
    \centering
    \includegraphics[width=\linewidth]{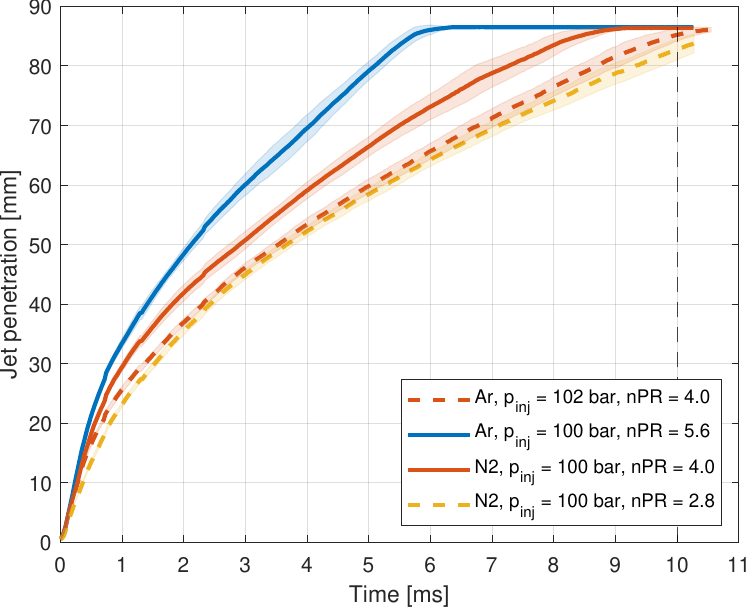}
    \caption{Hydrogen injections into argon and nitrogen when the density is matched at $\mathrm{p_a=\SI{25}{bar}}$ and $\mathrm{nPR = 4}$ (red lines) for either argon or nitrogen. The blue and yellow lines indicate the matched density by compensating $p_\mathrm{a}$ to argon (dashed line pair) or nitrogen (solid line pair).}
    \label{fig:densitymatch}
\end{figure}

While the injected mass and discharge coefficient are unaffected by the different ambient gases, jet penetration is not. The models of Tsujimura et al.\ and Ouellette \cite{tsujimura,Ouellette_1996} scale jet penetration inversely with the fourth root of chamber density ($S(t)\mathrm{\propto \rho_a^{-0.25}}$), indicating that jet penetration 'gradually' decreases with increasing $\rho_\mathrm{a}$. As $T_\mathrm{a}$ is constant at room temperature in the shown measurements, chamber density is only affected by gas composition and ambient gas pressure. Additional experiments were performed to match the density between the nitrogen and argon ambient by compensating the chamber pressure by $\mathrm{\frac{\rho_\mathrm{ar}}{\rho_\mathrm{N_2}} = 1.41}$ in \autoref{fig:densitymatch}, for both argon and for nitrogen at $p_\mathrm{a} =$ \SI{25}{bar}. Jet penetration is similar for conditions with lower pressure ratios (dashed lines, argon at \SI{25}{bar}), but deviates substantially at higher nPR conditions (solid lines, nitrogen at \SI{25}{bar}), not showing the same trend as earlier models, which is further investigated in \autoref{sub:emp}.

\subsection{Empirical jet penetration relationship}
\label{sub:emp}

\begin{figure}
    \centering
    \includegraphics[width=\linewidth]{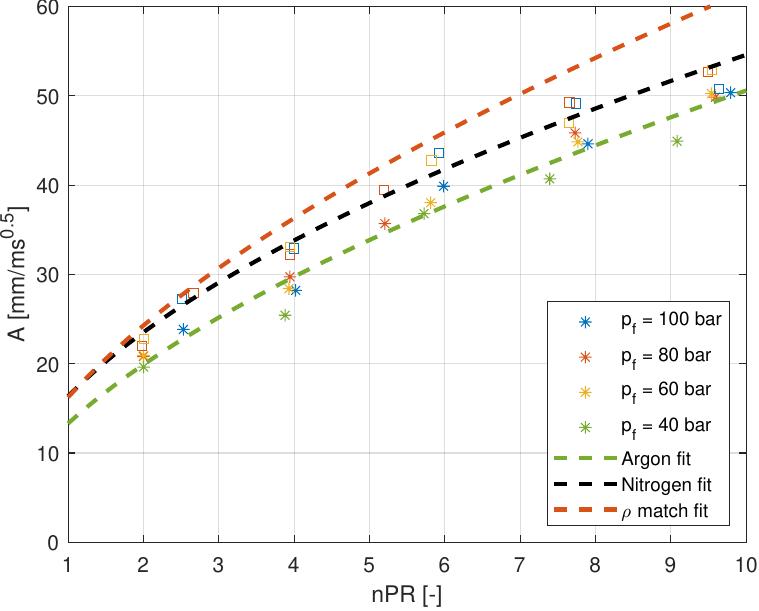}
    \caption{Fitted relationship between parameter $A$ (see \autoref{eq:jetpen_generic_formula}) and the pressure ratio for both argon (star symbols) and nitrogen (square symbols) environments. The computed fitting parameter from the experiments is distinguished for different injection pressures. The dashed lines show $A$ fitted to $\mathrm{nPR}$ using \autoref{eq:argonmaxmodel} (green), \autoref{eq:n2maxmodel} (black) or by including the linear density match $\mathrm{(\frac{\rho_{ar}}{\rho_{N_2}})^{0.58}}$ into $A$ of \autoref{eq:argonmaxmodel} (red).}
    \label{fig:jetpenmodel_A_fit}
\end{figure}

With a lacking match between the measured jet penetration results and the semi-empirical models of Ouellette \cite{Ouellette_1996} and Tsujimura \cite{tsujimura}, an empirical relationship for the jet penetration was derived in this work. In \autoref{fig:jepenmodel_generic} it was already shown that, starting approximately \SI{10}{mm} from the injector orifice, the jet penetration follows the constant rate of entrainment assumption closely. It is also clear that parameter $A$ changes with measurement conditions, while parameter $B$ is added to compensate for the opening stage of the injector and the buildup of flow. As constant $A$ is computed from the measurement data in \autoref{fig:jepenmodel_generic}, measurement conditions can be used to fit the computed constants in \autoref{eq:jetpen_generic_formula}. Given that nPR is the main parameter that affects the jet penetration, there was no trend found when comparing against injector timing or fuel and chamber pressure individually. For the opening offset $B$, no clear relationship was found for any of the studied variables, so it was chosen to take the mean value of the computed $B$ of the actual measurements in a corresponding ambient gas.  For the argon ambient this yields the following relationship:
\begin{equation}
S(t) = [13.31 \cdot nPR^{0.58}]\cdot  t^{1/2}-4.38,
    \label{eq:argonmaxmodel}
\end{equation}
while for nitrogen environments this amounted to:
\begin{equation}
S(t) = [16.37 \cdot nPR^{0.52}]\cdot  t^{1/2}-3.65,
    \label{eq:n2maxmodel}
\end{equation}
where $S(t)$ is in millimeter and $t$ in millisecond.

\begin{figure}
\centering
    \includegraphics[width=\linewidth]{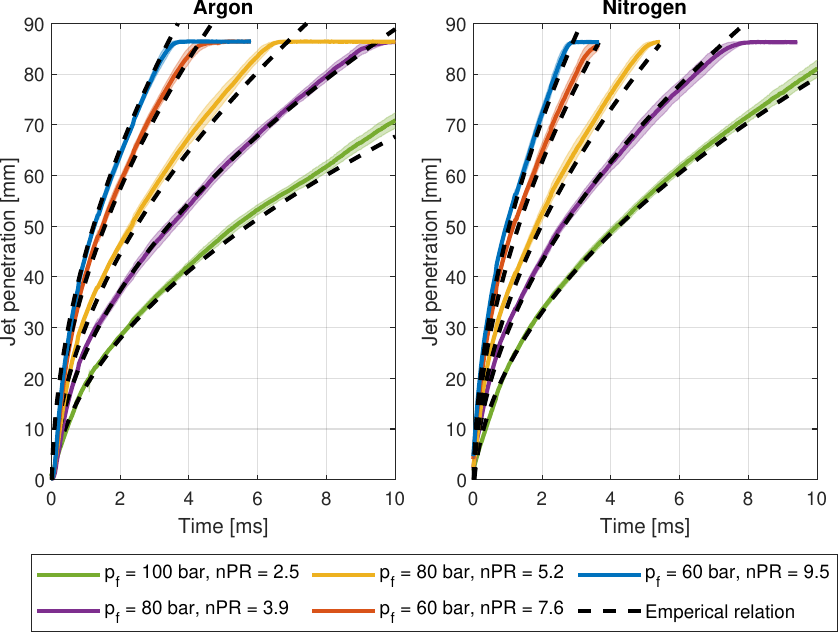}
    \caption{Empirical jet penetration results (black dashed) from \autoref{eq:argonmaxmodel} and \autoref{eq:n2maxmodel} compared to the experimental results.}
    \label{fig:jetpenmodel_maxfit}
\end{figure}

\autoref{fig:jetpenmodel_maxfit} shows the resulting penetration curves (dashed black lines) are mostly within the 95\% confidence interval of the measurement data. As shown in \autoref{fig:jetpenmodel_A_fit}, the penetration curves with $\mathrm{nPR \approx 6}$ deviates the most, falling just outside the confidence interval after \SI{70}{mm} penetration. It should be noted that the empirical constants $A$ and  $B$ are approximately 1.2 times bigger and smaller, respectively, for a nitrogen ambient compared to argon. This indicates faster penetration and reduced opening compensation, which would fall in line with the density ratio ${(\rho_\mathrm{Ar}/\rho_\mathrm{N_2})^{0.58}}$. The dashed red line in \autoref{fig:jetpenmodel_A_fit} is constructed by multiplying parameter $A$ for the argon measurements with this density ratio. However, that density ratio yielded less accurate jet penetration predictions at higher nPR's than \autoref{eq:n2maxmodel} due to the difference in the exponent (0.52 vs 0.58). As density rises linearly with pressure when assuming ideal gas conditions, this is also in agreement with the linear density match results shown in \autoref{fig:densitymatch}.

\section{Conclusions}
This work shows that a modified GDI injector can be suitable for controlling the injection of hydrogen into high-pressurized room-temperature ambient conditions at combustion engine-appropriate time scales.

Injected mass and discharge coefficients were measured and found to be in line with idealized choked flow theory, showing linear behavior for fuel pressure and injection duration, while there seems to be no dependency on the downstream (chamber) pressure. The injected mass is relatively low for combustion engine usage, but with a discharge coefficient below 0.5 there is potential to significantly increase the mass flow for the \SI{0.65}{mm} hole size, while a (conventional) multiple hole nozzle could further improve the mass flow of the injector.

The high-speed Schlieren measurements show adequate sensitivity for the low-density jets (compared to conventional liquid sprays), which allows for accurate post-processing of jet penetration and angle. Jet penetration is faster for higher pressure ratios and when comparing the nitrogen and argon ambient, mean jet penetration is always faster in nitrogen when all other variables are matched. When matching the density of two different ambient gases by linearly compensating the chamber pressure at the same ambient temperature, jet penetration still differed, the more so when the pressure ratio is increased.

Under the assumption of a constant rate of entrainment, an empirical jet penetration relationship was established based on the chosen pressure ratio, showing good agreement with the experimental results. Compared to continuous momentum jet models found in literature thus far, the jet angle also seems dictated by the pressure ratio in this work. Jet cone angle does not follow a clear trend on other studied variables. Therefore, the jet penetration and jet angle in this work can be considered as a result of the chosen pressure ratio instead of a parameter needed for a jet penetration model, which improves the understanding of $\mathrm{H_2}$ injection at high pressures which will contribute to the development of both the $\mathrm{H_2}$ ICE and Argon Power Cycle.  

\section*{Acknowledgements}
This work is part of the research program: “Argon Power Cycle” with project number 17868, which is partly financed by the Dutch Research Council (NWO). We would like to thank Noble Thermodynamic Systems (NTS) for supplying the modified Bosch GDI injector, with which the hydrogen jet experiments were performed. For the help and usage of the CT-scanner, we would like to express our gratitude to Marc van Maris and the Mechanics of Materials group at TU Eindhoven.

\section*{Nomenclature}
\centering
\begin{minipage}{.7\textwidth}

    \begin{table}[H]
\centering
\resizebox{\linewidth}{!}{

\begin{tabular}{lll}
\rowcolor[HTML]{343434} 
\multicolumn{1}{l|}{\cellcolor[HTML]{343434}{\color[HTML]{FFFFFF} Symbol}} & \multicolumn{1}{l|}{\cellcolor[HTML]{343434}{\color[HTML]{FFFFFF} Definition}} & {\color[HTML]{FFFFFF} unit}      \\ \hline
$\gamma$                                                                   & Specific heat ratio                                                            & -                                \\
\rowcolor[HTML]{EFEFEF} 
$V_\mathrm{c}$                                                             & Enclosed volume of vessel                                                      & $\mathrm{m^3}$                   \\
$p_\mathrm{i}$                                                             & Absolute pressure fuel                                                         & bar                              \\
\rowcolor[HTML]{EFEFEF} 
$T_\mathrm{i}$                                                             & Temperature                                                                    & K                                \\
$d_\mathrm{n}$                                                                      & Nozzle exit hole diameter                                                      & m                                \\
\rowcolor[HTML]{EFEFEF} 
$t_\mathrm{inj}$                                                                  & Injection duration                                                             & ms                               \\
nPR                                                                        & Pressure ratio                                                                 & -                                \\
\rowcolor[HTML]{EFEFEF} 
$\lambda$                                                                  & Wavelength                                                                     & nm                               \\
n                                                                          & Refractive index                                                               & -                                \\
\rowcolor[HTML]{EFEFEF} 
$\rho_\mathrm{i}$                                                          & Density                                                                        & $\mathrm{kg/m^3}$                \\
k                                                                          & Gladstone-Dale coefficient                                                     & -                                \\
\rowcolor[HTML]{EFEFEF} 
$\epsilon_\mathrm{i}$                                                      & Angular ray deflection                                                         & m                                \\
L                                                                          & Distance                                                                      & m                                \\
\rowcolor[HTML]{EFEFEF} 
I                                                                          & Intensity                                                                      & \# counts                        \\
S                                                                          & Jet penetration                                                                & mm                               \\
\rowcolor[HTML]{EFEFEF} 
t                                                                          & Time                                                                           & ms                               \\
A                                                                          & Fitting parameter                                                              & $\mathrm{\frac{mm}{\sqrt{ms}}}$          \\
\rowcolor[HTML]{EFEFEF} 
B                                                                          & Fitting parameter                                                              & mm                               \\
$\theta$                                                                   & Jet angle                                                                      & $^{\circ}$                       \\
\rowcolor[HTML]{EFEFEF} 
$dp_\mathrm{a}$                                                            & Mean pressure rise in vessel                                                   & bar                              \\
$m_\mathrm{inj}$                                                           & Injected mass                                                                  & mg                               \\
\rowcolor[HTML]{EFEFEF} 
R                                                                          & Specific gas constant                                                          & $\mathrm{\frac{kg}{kJ \cdot K}}$ \\
$\dot{m}_\mathrm{ideal}$                                                   & Idealized massflow                                                             & $\mathrm{\frac{mg}{ms}}$        \\
\rowcolor[HTML]{EFEFEF} 
A                                                                          & Area of exit hole                                                              & $\mathrm{m^2}$                   \\
$C_\mathrm{d}$                                                             & Discharge coefficient                                                          & -                                \\
\rowcolor[HTML]{EFEFEF} 
$n_\mathrm{a}$                                                             & Ambient species                                                                & -                               

\end{tabular}}
\end{table}

\end{minipage}

\bibliography{apc_SC}

\end{document}